\documentclass[twocolumn]{aastex62}

\newcommand{\G}{\textit{Gaia}}
\newcommand{\DR}{\textit{Gaia}~DR2}
\newcommand{\Sp}{\textit{Spitzer}}
\newcommand{\CW}{CatWISE}
\graphicspath{{./}{figures/}}

\received{April 22, 2019}
\revised{June 11, 2019}
\accepted{\today}
\submitjournal{ApJ}

\usepackage{bm}
\shorttitle{CWISEP J1935--1546}
\shortauthors{Marocco et al.}

\begin{document}

\title{CWISEP~J193518.59--154620.3: An Extremely Cold Brown Dwarf in the Solar Neighborhood Discovered with CatWISE}

\correspondingauthor{Federico Marocco}
\email{federico.marocco@jpl.nasa.gov}

\author[0000-0001-7519-1700]{Federico Marocco}\altaffiliation{NASA Postdoctoral Program Fellow}
\affiliation{Jet Propulsion Laboratory, California Institute of Technology, 4800 Oak Grove Dr., Pasadena, CA 91109, USA}
\affiliation{IPAC, Mail Code 100-22, Caltech, 1200 E. California Blvd., Pasadena, CA 91125, USA}

\author[0000-0001-7896-5791]{Dan Caselden}
\affiliation{Gigamon Applied Threat Research, 619 Western Avenue, Suite 200, Seattle, WA 98104, USA}

\author[0000-0002-1125-7384]{Aaron M. Meisner}\altaffiliation{Hubble Fellow}
\affiliation{National Optical Astronomy Observatory, 950 N. Cherry Ave., Tucson, AZ 85719, USA}

\author[0000-0003-4269-260X]{J. Davy Kirkpatrick}
\affiliation{IPAC, Mail Code 100-22, Caltech, 1200 E. California Blvd., Pasadena, CA 91125, USA}

\author[0000-0001-5058-1593]{Edward L. Wright}
\affiliation{Department of Physics and Astronomy, UCLA, 430 Portola Plaza, Box 951547, Los Angeles, CA 90095-1547, USA}

\author[0000-0001-6251-0573]{Jacqueline K. Faherty}
\affiliation{Department of Astrophysics, American Museum of Natural History, Central Park West at 79th Street, NY 10024, USA}

\author{Christopher R. Gelino}
\affiliation{IPAC, Mail Code 100-22, Caltech, 1200 E. California Blvd., Pasadena, CA 91125, USA}

\author{Peter R. M. Eisenhardt}
\affiliation{Jet Propulsion Laboratory, California Institute of Technology, 4800 Oak Grove Dr., Pasadena, CA 91109, USA}

\author{John W. Fowler}
\affiliation{230 Pacific St., Apt. 205, Santa Monica, CA 90405, USA}

\author[0000-0001-7780-3352]{Michael C. Cushing}
\affiliation{Department of Physics and Astronomy, University of Toledo, 2801 West Bancroft St., Toledo, OH 43606, USA}

\author[0000-0002-0077-2305]{Roc M. Cutri}
\affiliation{IPAC, Mail Code 100-22, Caltech, 1200 E. California Blvd., Pasadena, CA 91125, USA}

\author{Nelson Garcia}
\affiliation{IPAC, Mail Code 100-22, Caltech, 1200 E. California Blvd., Pasadena, CA 91125, USA}


\author[0000-0002-4939-734X]{Thomas H. Jarrett}
\affiliation{Department of Astronomy, University of Cape Town, Private Bag X3, Rondebosch, 7701, South Africa}

\author{Renata Koontz}
\affiliation{University of California, Riverside, 900 University Ave, Riverside, CA 92521, USA}
\affiliation{Jet Propulsion Laboratory, California Institute of Technology, 4800 Oak Grove Dr., Pasadena, CA 91109, USA}

\author{Amanda Mainzer}
\affiliation{Jet Propulsion Laboratory, California Institute of Technology, 4800 Oak Grove Dr., Pasadena, CA 91109, USA}

\author{Elijah J. Marchese}
\affiliation{University of California, Riverside, 900 University Ave, Riverside, CA 92521, USA}
\affiliation{Jet Propulsion Laboratory, California Institute of Technology, 4800 Oak Grove Dr., Pasadena, CA 91109, USA}

\author{Bahram Mobasher}
\affiliation{University of California, Riverside, 900 University Ave, Riverside, CA 92521, USA}

\author[0000-0002-5042-5088]{David J. Schlegel}
\affiliation{Lawrence Berkeley National Laboratory, Berkeley, CA 94720, USA}


\author[0000-0003-2686-9241]{Daniel Stern}
\affiliation{Jet Propulsion Laboratory, California Institute of Technology, 4800 Oak Grove Dr., Pasadena, CA 91109, USA}

\author[0000-0002-7064-5424]{Harry I. Teplitz}
\affiliation{IPAC, Mail Code 100-22, Caltech, 1200 E. California Blvd., Pasadena, CA 91125, USA}



\begin{abstract}

We present the discovery of an extremely cold, nearby brown dwarf in the solar neighborhood, found in the CatWISE catalog (Eisenhardt et al., in prep.). Photometric follow-up with \Sp\ reveals that the object, CWISEP~J193518.59--154620.3, has ch1--ch2\,=\,3.24$\pm$0.31\,mag, making it one of the reddest brown dwarfs known. Using the \Sp\ photometry and the polynomial relations from \citet{2019ApJS..240...19K} we estimate an effective temperature in the $\sim$270--360\,K range, and a distance estimate in the 5.6--10.9\,pc range. We combined the {\it WISE, NEOWISE}, and \Sp\ data to measure a proper motion of $\mu_\alpha \cos \delta = 337\pm69$\,mas\,yr$^{-1}$, $\mu_\delta = -50\pm97$\,mas\,yr$^{-1}$, which implies a relatively low tangential velocity in the range 7--22\,km\,s$^{-1}$. 

\end{abstract}



\section{Introduction} 
\label{sec:intro}
The census of objects in the solar neighborhood has been growing steadily in recent years \citep{2018AJ....155..265H}. The advent of large-area optical and near-infrared surveys (e.g. 2MASS, \citealt{2006AJ....131.1163S}; SDSS, \citealt{2000AJ....120.1579Y}; UKIDSS, \citealt{2007MNRAS.379.1599L}; VHS, \citealt{2013Msngr.154...35M}; PanSTARRS, \citealt{2016arXiv161205560C}; AllWISE, \citealt{2013wise.rept....1C}), and the recent \G\ second data release \citep{2018A&A...616A...1G}, have given us the opportunity to identify previously overlooked members of the 20\,pc sample. 

Despite recent discoveries of nearby ultracool dwarfs \citep{2018RNAAS...2a..33S,2018ApJ...868...44F,2018RNAAS...2b..50C,2018RNAAS...2d.205M}, the census of the coldest, lowest mass constituents of the solar neighborhood remains largely incomplete. \citet{2019ApJS..240...19K} found that the completeness limit for the T and Y dwarfs sample steeply declines as a function of effective temperature ($T_{\rm eff}$), from 19\,pc in the 900--1050\,K interval, down to 8\,pc in the 300--450\,K interval. At even lower $T_{\rm eff}$, only one object has been so far identified, WISE~J085510.83--071442.5 \citep[hereafter WISE~J0855--0714,][]{2014ApJ...786L..18L}, at a distance of 2.3\,pc. 

The paucity of objects in this temperature regime has prevented us from answering questions fundamental to astrophysics: how does star formation create objects of extremely low mass, and with what efficiency? Whereas the form of the mass function is well established for higher mass stars, it is far less constrained for the lowest mass stars and brown dwarfs. Objects of the lowest mass, including brown dwarfs, may, in fact, have several paths to creation depending upon their birth environment. Studies of star formation regions \citep[e.g.][]{2009A&A...508..823B} and nearby, young moving groups \citep{2016ApJS..225...10F,2017ApJ...837...95B} have shown that objects as low-mass as a few Jupiter masses ($M_{\rm Jup}$) can form in isolation. Older, isolated field objects with these masses therefore must exist, and will have had many Gyr to cool, making them cold analogs to planets in exosolar systems. Although establishing the diversity of low-mass star formation from cluster to cluster is important, observing more of these frigid, free-floating objects in the well-mixed field population will enable us to determine the frequency with which low-mass objects are formed across the age of the Galaxy.

Preliminary results from \citet{2019ApJS..240...19K} show that the cutoff mass for star formation, if there is one, must be lower than 10\,$M_{\rm Jup}$ since the model predictions assuming such a cutoff are already underpredicting the number of objects found. In order to test cutoffs of lower mass, we require more objects of extremely low temperature, akin to WISE~J0855--0714. 

As its name suggests, WISE~J0855--0714 was discovered using data from {\it WISE} \citep{2010AJ....140.1868W}, whose $W1$ (3.4\,$\mu$m) and $W2$ (4.6\,$\mu$m) bands are ideally placed to identify extremely cold brown dwarfs with their red $W1-W2$ colors (in contrast to the blue $W1-W2$ colors of stars) because the $W2$ band measures the peak of the spectral energy distribution while the $W1$ band lies in a region of strong methane absorption \citep{1997ApJ...491..856B}. WISE~J0855--0714 is relatively bright in $W2$ \citep[13.89$\pm$0.05\,mag,][]{2014ApJ...786L..18L}, and so, despite the ``statistics of one'', \citet{2014AJ....148...82W} estimated the 68\% confidence range for the number of ``0855-like'' objects in the existing AllWISE dataset to be 4--35, with a median of 15.

Finding more of these hidden solar neighbors is one of the goals of CatWISE, a NASA Astrophysics Data Analysis Program (ADAP) funded project combining data from the 2010 to 2016 phases of the {\it WISE} mission, to generate an all-sky photometric and astrometric catalog (Eisenhardt et al., in prep.).

Here we present CWISEP~J193518.59--154620.3 (hereafter CWISEP~J1935--1546), an extremely cold brown dwarf at $\sim$8\,pc discovered in the preliminary CatWISE catalog\footnote{CWISEP is the official designation for sources identified in the \CW\ Preliminary catalog, see Eisenhardt et al., in prep.}. Its $W1-W2$ color, and follow-up \Sp\ photometry, suggest CWISEP~J1935--1546 has an effective temperature comparable to that of WISE~J0855--0714, in the 270--360\,K range, making it one of the coldest brown dwarfs identified so far.

In Section~\ref{sec:catwise}, we briefly describe the \CW\ data processing and the preliminary catalog content; Section~\ref{sec:tsel} details the machine-learning-based procedure used to identify CWISEP~J1935--1546; in Section~\ref{sec:spitzer} we present our \Sp\ follow-up photometry, and in Section~\ref{sec:astrometry} we combine the \Sp\ data with the \textit{WISE} data to refine the motion measurement for this target. Finally, in Section~\ref{sec:analysis} we derive the basic properties for this cold new member of the solar neighborhood.

\section{CatWISE}
\label{sec:catwise}
\CW\ is an infrared photometric and astrometric catalog consisting of 900,849,014 sources over the entire sky selected from {\it WISE} and {\it NEOWISE} data collected from 2010 to 2016 at $W1$ and $W2$.

CatWISE adapted the AllWISE pipeline to work on the coadded WISE and NEOWISE images provided by unWISE \citep{2018AJ....156...69M,2018RNAAS...2d.202M}. A full description of CatWISE is provided in Eisenhardt et al. (in prep.), and of the AllWISE pipeline in \citet{2013wise.rept....1C} and \citet{2014ApJ...783..122K}. Here we summarize the steps relevant to the discovery of CWISEP~J1935--1546.

Source detection for the preliminary CatWISE catalog was performed using \textsc{MDET} \citep{2012PASA...29..269M}, which works simultaneously in $W1$ and $W2$. The full-depth unWISE coadds \citep{2018RNAAS...2d.202M} were resampled from 2048$\times$2048  (2.75$''$/pixel) format to the 4095$\times$4095 (1.375$''$/pixel) format used by \textsc{MDET} for \textit{WISE} source detection, using the Image Co-addition with Optional Resolution Enhancement software \citep[\textsc{ICORE},][]{2013ascl.soft02010M}, and an appropriate point spread function (PSF). The PSF interpolation kernel smooths the images, providing a matched filter for optimal detection of isolated point sources. The ``std'' unWISE images were used for uncertainties, as these provide the standard deviation at each coadd pixel of the individual \textit{WISE} exposures. The detection threshold was set at SNR\,=\,1.8, yielding a differential source reliability of 50\% based on deeper \Sp\ data from the S-COSMOS program \citep{2007ApJS..172...86S}.  

The \textsc{WPHOT} software package developed for the AllWISE pipeline \citep{2013wise.rept....1C}, was adapted to perform source photometry and astrometry for CatWISE. Two main changes are worth highlighting here:

(1) For AllWISE, \textsc{WPHOT} propagates each source position detected by \textsc{MDET} in the coadded images to individual exposures, solving for the least-square best-fit to the PSF, to determine source position and fluxes (hereafter ``stationary fit''). An alternative fit is also performed, allowing for linear motion of the source through the individual exposures (hereafter ``motion fit''). This solution provides, along with motion, an alternative measurement of position, propagated to a chosen reference epoch, as well as fluxes.

For a given inertial position in
each sky coverage, the $\sim$12 exposures which are combined in each unWISE epoch coadd \citep{2018AJ....156...69M} are obtained within less than two days. As a result, the position of sources beyond the solar system can be assumed to be fixed for each epoch. Therefore \CW\ used unWISE epoch coadd images in place of individual exposures when running \textsc{WPHOT}.

(2) The other significant modification to the AllWISE version of \textsc{WPHOT} involves the treatment of the PSF asymmetry. The \textit{WISE} PSF is asymmetric with respect to the scanning direction (which is along lines of ecliptic longitude). The scan direction is similar for all individual images in an epoch coadd, and a given inertial position is scanned in opposite directions every six months, i.e. in consecutive epochs (except very near an ecliptic pole). Since \textsc{WPHOT} was not designed to use a time-dependent PSF, \CW\ chose to measure source properties separately for the (typically four or more) epoch coadds in each of the two scan directions, and then merge the two results.

The Preliminary \CW\ Catalog is available at \url{catwise.github.io}, and will soon be available on the NASA/IPAC Infrared Science Archive\footnote{\url{https://irsa.ipac.caltech.edu/}} as well. From comparison to \Sp, the signal-to-noise-ratio\,=\,5 Vega magnitude limits are $W1=17.58$\,mag and $W2=16.43$\,mag (cf. $W1=16.90$\,mag and $W2=15.95$\,mag for AllWISE). From comparison to \DR\ \citep{2018A&A...616A...2L}, \CW\ measures motions to a 1$\sigma$ accuracy of 100 mas yr$^{-1}$ for sources $\sim$3\,mag fainter than those measured with a similar accuracy in AllWISE, and and achieves motion measurement accuracy that is 10 times better at $W2 = 15$\,mag (Eisenhardt et al., in prep.), as expected given the longer time baseline afforded by the combined data.

\section{Target selection}
\label{sec:tsel}
CWISEP~J1935--1546 was found as part of our larger effort to identify and characterize very cold brown dwarfs using \CW. The search was conducted using the \textsc{Python} package \textit{XGBoost}\footnote{\url{https://xgboost.readthedocs.io/en/latest/}} \citep{Chen:2016:XST:2939672.2939785}, which implements machine learning algorithms under the gradient boosting framework. 

In general applications of supervised learning with \textit{XGBoost}, one trains and evaluates an \textit{XGBoost} model with a set of previously classified samples. This ground truth dataset is used throughout the development process to find a model with effective features and hyperparameters. Our application followed this general supervised learning paradigm. 

We evaluated multiple \textit{XGBoost} classifiers in target selection. CWISEP~J1935--1546 was selected by our classifier that targets faint and red objects. More specifically, the classifier is restricted to training with and classifying point sources with $W2 > 14$\,mag and $W1-W2 > 1$\,mag, or within $3\sigma$ of those limits, therefore consistent with T or Y spectral type \citep[see e.g.][]{2011ApJS..197...19K}.

To build our training set, since the prevalence of our target class (herein ``positive'') is so low, and our compensatory sample weights for the remainder of the data set (herein ``negative'') are also low by consequence, we manually classified samples for the positive class versus randomly sampled for the negative class. We took confirmed objects from the literature, and motion-confirmed objects from our candidate lists, and cross-matched them against \CW\ to obtain their \CW\ data. We removed/corrected mismatches until we were confident that the remaining training set held an insignificant number of mismatched training objects. This set, consisting of $\approx$200 objects, became our positive class. 

We carefully selected sample weights to achieve robust classification. Firstly, we weighted samples of the positive and negative class as the inverse proportion of the total number of objects in each class within the training data. Consequently, samples belonging to the low population positive class received higher weight, and those belonging to the high population negative class received lower
weight, so that

\begin{equation}
    \sum_{i=0}^{n_{neg}} w_{neg,i} = \sum_{i=0}^{n_{pos}} w_{pos,i}
\end{equation}

\noindent where $w_{neg,i}$, $w_{pos,i}$ are the weight for a single member of the negative and positive class, respectively, and $n_{neg}$, $n_{pos}$ are the total number of objects in the negative and positive class.

Next, we distributed weights within the positive class by $W2$ magnitude, creating an even distribution of total weights per given 0.5 bin of $W2$ magnitude. That is, our training data contained fewer faint $W2$ magnitude positive class members, so their weights were proportionally higher than bright $W2$ magnitude positive class members. In practice

\begin{equation}
        \sum_{i=0}^{n_{pos,W2}} w_{pos,W2,i} = \sum_{i=0}^{n_{pos,W2+0.5}} w_{pos,W2+0.5,i}
\end{equation}

\noindent where $w_{pos,W2,i}$, $w_{pos,W2+0.5,i}$ are the weight for a single member of the positive class in a given 0.5 magnitude bin (e.g. $14.5 \leq W2 < 15.0$\,mag) and in the adjacent magnitude bin, respectively, while $n_{pos,W2}$, $n_{pos,W2+0.5}$ are the total number of objects in the two bins in question.

Machine-learning classifiers are defined by two different sets of parameters -- model parameters and hyperparameters (also referred to as tuning parameters). Model parameters are estimated by the machine learning algorithm itself, from the data, as part of the learning process. Hyperparameters on the other hand cannot be estimated directly from the data, as they regulate the learning process itself \citep[see e.g.][]{kuhn2013applied}. Examples of hyperparameters are the k in k-nearest neighbor interpolation, or the learning rate for training. Finding the optimal set of hyperparameters is itself a complex problem, and several approaches have been adopted \citep[see][and references therein]{2015arXiv150202127C}.
 
We searched with {\it Scikit-learn}'s randomized cross-validation function for optimized hyperparameters \citep[{\it RandomizedSearchCV};][]{pedregosa2011scikit}. The function takes in a model, the training set, a selection of hyperparameters, and distributions from which to draw their guesses. It then picks hyperparameter values from the provided distributions, and trains and tests the model, searching for the values that optimize the model performance. However, experimentally we found that a low learning rate of 0.0135, manually enforced outside of the parameter search, lead to the greatest reduction of the classification error rate. The classification error rate is defined as $n_{\rm wrong}/n_{\rm tot}$, where $n_{\rm wrong}$ is the number of misclassified objects, and $n_{\rm tot}$ is the total number of classified objects \citep[see e.g.][Chapter 4.2]{tan2018introduction}. 

After first training each \textit{XGBoost} classifier with our initial training set, we applied it to the entire \CW\ catalog, and selected the objects (usually between 10,000 and 25,000) with the highest predicted probability membership in the positive class. We then visually inspected each object, using available optical, near- and mid-infrared images (taken from DSS, 2MASS, UKIDSS, PanSTARRS, and AllWISE) and the online image blinking/visualization tool WiseView\footnote{\url{http://byw.tools/wiseview}} \citep{2018ascl.soft06004C}. Objects confirmed to be real, with $W1-W2$ color visually consistent with $W1-W2 > 1$\,mag, and with visible motion (confirming they are nearby), were added to the positive class. Common false positives included objects that were found to be unflagged artifacts, variable sources leading to spurious motion measurements, and partly blended objects with contaminated photometry and/or astrometry.

We then iterated by re-training the classifier on the full training data, and applied the re-trained classifier to the entire catalog to select another batch of high probability positive class entries. Periodically, we would validate both the manually labelled and randomly sampled training data to remove mislabelled objects. We would do this by performing various train-test splits and visually inspecting negatively labelled entries that had the highest probability (among such entries) of belonging to the positive class, as well as the converse case. 

The selection yielded an initial sample of 131 late-T and Y dwarf candidates. After further visual inspection, we prioritized 32 with either no detection or a marginal detection in W1 (hinting at an extremely low temperature) and visible motion (hinting at their proximity). These are being followed-up through our \Sp\ campaign (see Section~\ref{sec:spitzer}) to obtain ch1 ($3.6\,\mu$m) and ch2 ($4.5\,\mu$m) photometry to confirm/refute their nature and estimate effective temperature and photometric distance. CWISEP~J1935--1546 is the reddest among the objects followed-up so far, with $W2 = 15.926\pm0.085$\,mag, $W1-W2 = 2.58\pm0.37$\,mag , and ch1--ch2 = 3.24$\pm$0.31\,mag (see next section).

\section{{\it Spitzer} follow-up}
\label{sec:spitzer}
\Sp\ observations were taken as part of program 14034 (Meisner, PI). Seven exposures of 30\,s were taken in each band, and these exposures were dithered using a random dither pattern of medium scale. The number of individual exposures was chosen so that we would obtain a 5$\sigma$ detection at ch1--ch2\,=\,2.75\,mag. 

Our target is very faint in the ch1 mosaic, and to measure it we had to lower the SNR for detection from 5 (the default value in MOPEX/APEX) down to 2. For the aperture photometry, we used an aperture with a radius of 4 pixels ({\it aperture1} in the MOPEX output files) and a sky annulus with a 24-to-40-pixels radius. For the PRF-fit photometry, we used a set of warm PRFs built by Jim Ingalls (see \citealt{2019ApJS..240...19K}), which are very similar to the warm mission PRFs developed by \cite{2012SPIE.8442E..39H}, that are available on the IRSA website\footnote{See \url{http://irsa.ipac.caltech.edu/data/SPITZER/docs/irac/calibrationfiles/}.}. For the aperture photometry, the resulting raw fluxes were multiplied by the aperture corrections recommended in Table 4.7 of the IRAC Instrument Handbook -- 1.208 for ch1 and 1.221 for ch2 -- to obtain the flux in units of $\mu$Jy; for PRF-fit photometry, the resulting raw fluxes were divided by the correction factor recommended in Table C.1 of the IRAC Instrument Handbook -- 1.021 for ch1 and 1.012 for ch2. These aperture and PRF-fit fluxes were then converted from $\mu$Jy to magnitudes using the flux zero points in the Handbook's Table 4.1 (280.9$\pm$4.1 Jy in ch1 and 179.7$\pm$2.6 Jy in ch2), propagating the uncertainty in zero point and flux into the final measurement error. This final photometry is given in Table~\ref{tab:photometry}. The ch1 detection reported here corresponds to a SNR of 3.6.

\begin{figure*}
    \centering
    \includegraphics[width=\textwidth, trim={3cm 0 3cm 0}, clip]{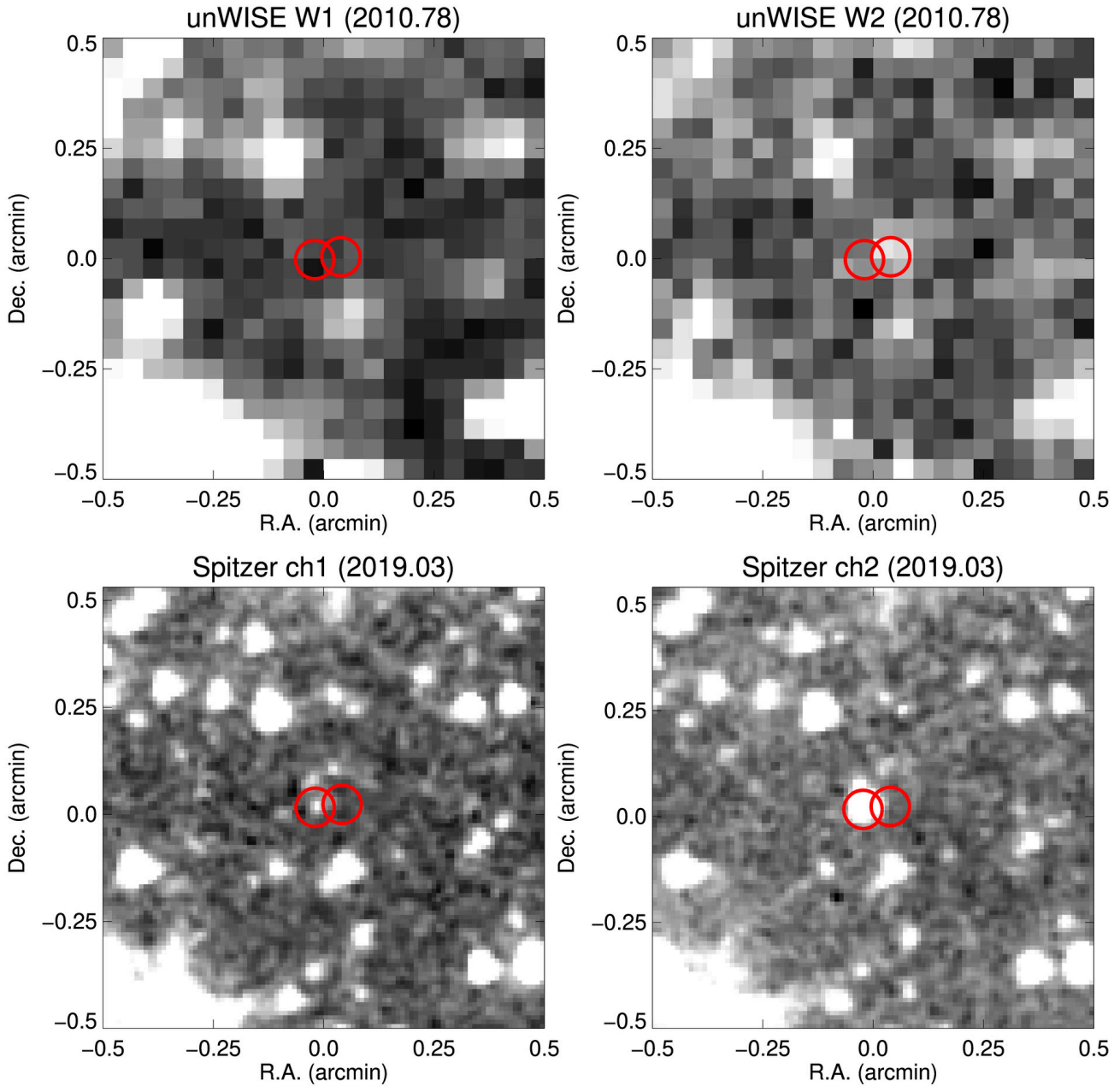}
    \caption{$1\times1$ arcmin cutouts from the unWISE W1 and W2 epoch coadds (top left and right), and the \Sp\ ch1 and ch2 mosaic (bottom left and right), centered around CWISEP~J1935--1546. Red circles mark its position at the two epochs shown.}
    \label{fig:images}
\end{figure*}

CWISEP~J1935--1546 has ch1--ch2 = 3.24$\pm$0.31 mag (PRF; the aperture color is 3.46$\pm$0.26), overlapping with WISE~J0855--0714 (3.55$\pm$0.07\,mag), and similar to the second reddest brown dwarf known, WISE~J035000.32-565830.2 \citep[3.25$\pm$0.10 mag;][]{2012ApJ...753..156K}. Figure~\ref{fig:images} shows the unWISE $W1$ and $W2$ coadds, and our \Sp\ ch1 and ch2 mosaics.

\begin{table}
    \centering
    \caption{\CW\ and \Sp\ photometry for CWISEP~J1935--1546.}
    \label{tab:photometry}
    \begin{tabular}{l c c}
    \hline
    Band & Magnitude & Notes \\
    \hline
    CatWISE W1 & 18.509$\pm$0.359 & stationary fit ({\it w1mpro}) \\
    CatWISE W2 & 15.926$\pm$0.085 & stationary fit ({\it w2mpro}) \\
    CatWISE W1 & 18.534$\pm$0.396 & motion fit ({\it w1mpro\_pm}) \\
    CatWISE W2 & 15.852$\pm$0.079 & motion fit ({\it w2mpro\_pm}) \\
    \Sp\ ch1 & 19.089$\pm$0.262 & aperture \\
    \Sp\ ch2 & 15.633$\pm$0.018 & aperture \\
    \Sp\ ch1 & 18.892$\pm$0.314 & PRF fit \\
    \Sp\ ch2 & 15.647$\pm$0.023 & PRF fit \\
    \hline
    \end{tabular}
\end{table}

\section{Astrometry}
\label{sec:astrometry}
The CWISEP~J1935--1546 \CW\ measured motion is $\mu_\alpha \cos \delta = 400\pm100$ mas yr$^{-1}$, $\mu_\delta = -90\pm120$ mas yr$^{-1}$. However, with the aid of our \Sp\ follow-up observation, we have obtained a better measurement of the target's motion. 

We first re-registered the unWISE epoch coadds to the \G\ astrometric frame. We extracted sources from the individual epoch coadds using the \CW\ pipeline. Because that was run on individual epochs, no ``motion fit'' was possible, and therefore the positions used are those resulting from the ``stationary fit'' (see Section~\ref{sec:catwise}).

We then selected a sample of bright reference stars to be used for re-registration. We retained only stars with $\sigma_\alpha$, $\sigma_\delta < 0.1''$, $W1\,>\,8.1$\,mag and $W2\,>\,6.7$\,mag \citep[the saturation limits for \textit{WISE};][]{2012wise.rept....1C}. 
After these cuts were applied, our initial re-registration set consisted of more than 10,000 stars at each epoch.

We cross-matched our re-registration set to \DR. We used \DR\ astrometry to correct the positions of all of the \G\ stars in the \CW\ field to the \CW\ epoch in question, and matched them to the re-registration set using a 2.75$''$ radius (corresponding to one unWISE pixel), retaining only the closest matching \G\ source for each re-registration star. We typically found $\sim$700 \G\ stars at each epoch. 

The re-registration stars were used to fit both a 6th order and a 12th order transformation between each epoch coadd and \DR, using our own \textsc{IDL} code. The fit included a $3\sigma$ clipping iteration, removing re-registration stars whose re-registered position after the first fit iteration was off by more than three times the formal errors of the fit from their \DR\ position. The use of a 12th order transformation did not significantly reduce the residuals of the fit, and therefore we adopted the 6th order transformation.

To determine reliable uncertainties on the coordinates of CWISEP~J1935--1546 at each epoch, we examined the dispersion between the re-registered positions of all stars in the field (without any restriction on their positional accuracy), and their \DR\ positions. We found that at $W2 \sim 16$\,mag (the brightness of CWISEP~J1935--1546) the formal uncertainties reported by \textsc{WPHOT} underestimate by a factor of $\sim$1.6 the observed dispersion. We therefore multiply the formal uncertainties by that factor. \CW\ positions and our adopted uncertainties are listed in Table~\ref{tab:astrometry}.

For the re-registration of the \Sp\ ch2 mosaic, and the measurement of CWISEP~J1935--1546 at that epoch, we adopted the same method described in \citet{2019ApJS..240...19K}.

Finally, we performed both a linear fit and a 5-parameter astrometric fit (position + proper motion + parallax motion) to the \CW\ and \Sp\ ch2 positions of CWISEP~J1935--1546 as a function of time (listed in Table~\ref{tab:astrometry}). The target falls below the signal-to-noise-ratio threshold for detection by the \CW\ pipeline (SNR = 1.8) in two of the nine {\it WISE} plus {\it NEOWISE} epochs. 

\begin{table}
    \centering
    \caption{Measured positions for CWISEP~J1935--1546. The 2019 position was obtained through our \Sp\ follow-up. The target is undetected in two of the nine unWISE epoch coadds.}
    \label{tab:astrometry}
    \begin{tabular}{c c c c c}
    \hline
    R.A. & $\sigma_{\rm R.A.}$ & Dec. & $\sigma_{\rm Dec.}$ & Epoch \\
    (deg) & (arcsec) & (deg) & (arcsec) & \\
    \hline
    293.8267517 & 0.9 & -15.7722349 & 1.4 & 2010.2842 \\
    293.8270569 & 1.0 & -15.7723198 & 1.3 & 2010.7764 \\
    293.8275452 & 1.2 & -15.7720318 & 1.1 & 2015.2857 \\
    293.8275146 & 1.1 & -15.7723351 & 1.4 & 2015.7695 \\
    293.8275452 & 1.1 & -15.7723131 & 1.3 & 2016.7525 \\
    293.8276062 & 1.4 & -15.7726307 & 1.4 & 2017.2799 \\
    293.8276978 & 1.1 & -15.7723455 & 1.2 & 2017.7436 \\
    293.8277930 & 0.01 & -15.7723462 & 0.01 & 2019.0267 \\
    \hline
    \end{tabular}
\end{table}

\begin{figure}
    \centering
    \includegraphics[width=0.49\textwidth, trim={2cm 3cm 1cm 3cm}, clip]{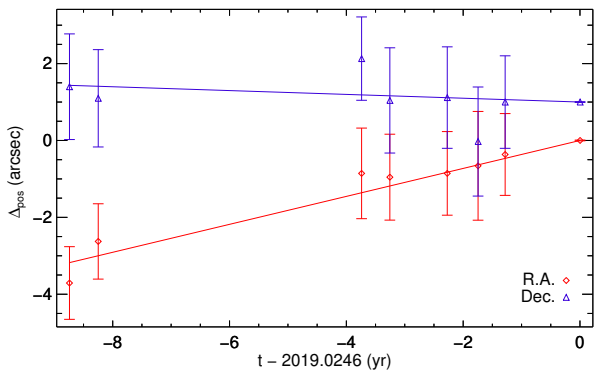}
    \caption{A linear fit to the coordinates of CWISEP~J1935--1546 as a function of time. R.A. (red diamonds), Dec. (blue triangles), and time are relative to their variance-weighted mean.}
    \label{fig:pm}
\end{figure}

The linear fit yielded $\mu_\alpha \cos \delta = 337\pm69$\,mas\,yr$^{-1}$, $\mu_\delta = -50\pm97$\,mas\,yr$^{-1}$, while the 5-parameter astrometric fit yielded $\mu_\alpha \cos \delta = 341\pm90$\,mas\,yr$^{-1}$, $\mu_\delta = -36\pm113$\,mas\,yr$^{-1}$, and a trigonometric parallax of $-100\pm440$\,mas. The negative parallax is clearly unphysical and not statistically significant. We therefore adopt the values from the linear fit. The results of the linear fit are presented in Figure~\ref{fig:pm}.

\section{Analysis}
\label{sec:analysis}
\begin{figure}
    \centering
    \includegraphics[width=0.49\textwidth, trim={2cm 3cm 1cm 3cm}, clip]{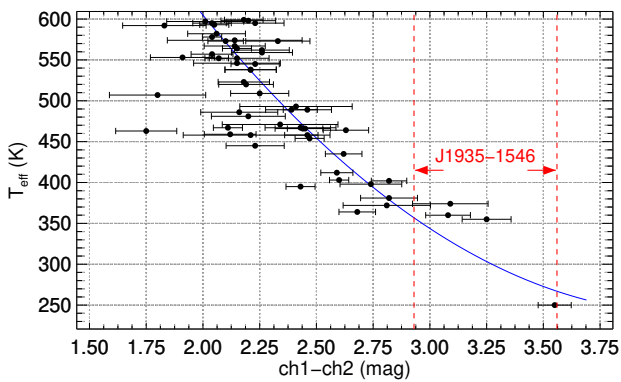}
    \caption{$T_{\rm eff}$ as a function of \Sp\ ch1--ch2 colors for nearby late-T and Y dwarfs. Black points are objects taken from \citet[Table 8]{2019ApJS..240...19K}. The red dashed lines encompass the $1\sigma$ color range for CWISEP~J1935--1546. Overplotted in blue is the polynomial relation presented in \citet{2019ApJS..240...19K}.}
    \label{fig:cmd}
\end{figure}

The ch1--ch2 color for CWISEP~J1935--1546 is comparable to that of the coldest brown dwarf known, WISE~J0855--0714. In Figure~\ref{fig:cmd} we show $T_{\rm eff}$ as a function of \Sp\ ch1--ch2 color for a sample of known late-T and Y dwarfs from the literature \citep[see][and references therein]{2019ApJS..240...19K}. 

We can use our \Sp\ photometry and the recent polynomial relations presented in \citet{2019ApJS..240...19K} to derive an effective temperature estimate and photometric distance for CWISEP~J1935--1546. Using the ch1--ch2 color to absolute ch2 magnitude relation, we obtain a $1\sigma$ distance estimate of 5.6--10.9\,pc. Given this estimated distance and the proper motion measured here, CWISEP~J1935–1546 has an estimated tangential velocity in the range 7--22\,km\,s$^{-1}$, consistent with the tangential velocity distribution for L, T and Y dwarfs in the solar neighborhood from \citet{2019MNRAS.tmp..664S} and \citet{2019ApJS..240...19K}.

The ch1--ch2 colour to $T_{\rm eff}$  relation from \citet{2019ApJS..240...19K} indicates a $T_{\rm eff}$ in the range $\sim$270--360\,K, which would make CWISEP~J1935--1546 one of the coldest brown dwarfs discovered so far (see Figure~\ref{fig:cmd}).

Given the temperature derived above, we can estimate a mass for this object using the BT-Settl models \citep{2012EAS....57....3A,2013MmSAI..84.1053A}. If we assume CWISEP~J1935--1546 is a field object, with age in the $\sim$500\,Myr -- 13\,Gyr range, it would have a mass in the range 2--20\,$M_{\rm Jup}$. We can narrow down the age and mass range by taking into account the fact that the tangential velocity estimated here is consistent with the population of nearby ultracool dwarfs, whose age is in the range $\sim$1.5--6.5\,Gyr \citep[see e.g.][and references therein]{2018PASP..130f4402W}. In this age range, CWISEP~J1935--1546 would have a mass between 3 and 14\,$M_{\rm Jup}$. 

No other photometry is currently available for this object, as it is well below the detection threshold for existing optical and near-infrared surveys. The position of CWISEP~J1935--1546 is covered by VHS and PanSTARRS, but the target is undetected in both, as well as in the $W3$ and $W4$ images from AllWISE. Given the $T_{\rm eff}$ and distance ranges estimated above, the expected $H$ magnitude would be 23.7--25.1\,mag, a depth prohibitive for most ground-based facilities, particularly for spectroscopy. Spectroscopic characterization for this extremely cold object will necessarily have to wait for the launch of {\it JWST}.

The discovery of CWISEP~J1935--1546 starts to bridge the existing gap between known warmer Y dwarfs and the extremely cold WISE~J0855--0714. \CW, as well as ``Backyard Worlds: Planet 9'' \citep[a NASA-funded citizen science project;][]{2017ApJ...841L..19K} are now fully exploiting the potential of the {\it WISE} and {\it NEOWISE} data set to uncover more of these frigid, free-floating planetary mass objects. CWISEP~J1935--1546 is part of a larger sample of discoveries by these two highly complementary projects, and joint observing campaigns are now underway to fully characterize this compelling population. By populating this region of parameter space we can not only put strong observational constraints on the mass function for extremely low mass objects, but also understand the processes that shape such cold, planet-like atmospheres.

\acknowledgments
This research was partly carried out at the Jet Propulsion Laboratory, California Institute of Technology, under a contract with NASA. 

FM is supported by an appointment to the NASA Postdoctoral Program at the Jet Propulsion Laboratory, administered by Universities Space Research Association under contract with NASA.

AMM acknowledges support from Hubble Fellowship HST-HF2-51415.001-A.

CatWISE is funded by NASA under Proposal No. 16-ADAP16-0077 issued through the Astrophysics Data Analysis Program, and uses data from the NASA-funded WISE and NEOWISE projects.

This work is based in part on observations made with the Spitzer Space Telescope, which is operated by the Jet Propulsion Laboratory, California Institute of Technology under a contract with NASA.

\bibliographystyle{aasjournal}
\bibliography{refs.bib}

\begin{thebibliography}{}
\expandafter\ifx\csname natexlab\endcsname\relax\def\natexlab#1{#1}\fi
\providecommand{\url}[1]{\href{#1}{#1}}

\bibitem[{{Allard} {et~al.}(2013){Allard}, {Homeier}, \&
  {Freytag}}]{2013MmSAI..84.1053A}
{Allard}, F., {Homeier}, D., \& {Freytag}, B. 2013, \memsai, 84, 1053

\bibitem[{{Allard} {et~al.}(2012){Allard}, {Homeier}, {Freytag}, \&
  {Sharp}}]{2012EAS....57....3A}
{Allard}, F., {Homeier}, D., {Freytag}, B., \& {Sharp}, C.~M. 2012, in EAS
  Publications Series, Vol.~57, EAS Publications Series, ed. C.~{Reyl{\'e}},
  C.~{Charbonnel}, \& M.~{Schultheis}, 3--43

\bibitem[{{Best} {et~al.}(2017){Best}, {Liu}, {Magnier}, {Bowler}, {Aller},
  {Zhang}, {Kotson}, {Burgett}, {Chambers}, {Draper}, {Flewelling}, {Hodapp},
  {Kaiser}, {Metcalfe}, {Wainscoat}, \& {Waters}}]{2017ApJ...837...95B}
{Best}, W.~M.~J., {Liu}, M.~C., {Magnier}, E.~A., {et~al.} 2017, \apj, 837, 95

\bibitem[{{Burgess} {et~al.}(2009){Burgess}, {Moraux}, {Bouvier}, {Marmo},
  {Albert}, \& {Bouy}}]{2009A&A...508..823B}
{Burgess}, A.~S.~M., {Moraux}, E., {Bouvier}, J., {et~al.} 2009, \aap, 508, 823

\bibitem[{{Burrows} {et~al.}(1997){Burrows}, {Marley}, {Hubbard}, {Lunine},
  {Guillot}, {Saumon}, {Freedman}, {Sudarsky}, \&
  {Sharp}}]{1997ApJ...491..856B}
{Burrows}, A., {Marley}, M., {Hubbard}, W.~B., {et~al.} 1997, \apj, 491, 856

\bibitem[{{Caselden} {et~al.}(2018){Caselden}, {Westin}, {Meisner}, {Kuchner},
  \& {Colin}}]{2018ascl.soft06004C}
{Caselden}, D., {Westin}, III, P., {Meisner}, A., {Kuchner}, M., \& {Colin}, G.
  2018, {WiseView: Visualizing motion and variability of faint WISE sources},
  Astrophysics Source Code Library, , , ascl:1806.004

\bibitem[{{Chambers} {et~al.}(2016){Chambers}, {Magnier}, {Metcalfe},
  {Flewelling}, {Huber}, {Waters}, {Denneau}, {Draper}, {Farrow}, {Finkbeiner},
  {Holmberg}, {Koppenhoefer}, {Price}, {Saglia}, {Schlafly}, {Smartt},
  {Sweeney}, {Wainscoat}, {Burgett}, {Grav}, {Heasley}, {Hodapp}, {Jedicke},
  {Kaiser}, {Kudritzki}, {Luppino}, {Lupton}, {Monet}, {Morgan}, {Onaka},
  {Stubbs}, {Tonry}, {Banados}, {Bell}, {Bender}, {Bernard}, {Botticella},
  {Casertano}, {Chastel}, {Chen}, {Chen}, {Cole}, {Deacon}, {Frenk},
  {Fitzsimmons}, {Gezari}, {Goessl}, {Goggia}, {Goldman}, {Grebel}, {Hambly},
  {Hasinger}, {Heavens}, {Heckman}, {Henderson}, {Henning}, {Holman}, {Hopp},
  {Ip}, {Isani}, {Keyes}, {Koekemoer}, {Kotak}, {Long}, {Lucey}, {Liu},
  {Martin}, {McLean}, {Morganson}, {Murphy}, {Nieto-Santisteban}, {Norberg},
  {Peacock}, {Pier}, {Postman}, {Primak}, {Rae}, {Rest}, {Riess}, {Riffeser},
  {Rix}, {Roser}, {Schilbach}, {Schultz}, {Scolnic}, {Szalay}, {Seitz},
  {Shiao}, {Small}, {Smith}, {Soderblom}, {Taylor}, {Thakar}, {Thiel},
  {Thilker}, {Urata}, {Valenti}, {Walter}, {Watters}, {Werner}, {White},
  {Wood-Vasey}, \& {Wyse}}]{2016arXiv161205560C}
{Chambers}, K.~C., {Magnier}, E.~A., {Metcalfe}, N., {et~al.} 2016, ArXiv
  e-prints, arXiv:1612.05560

\bibitem[{Chen \& Guestrin(2016)}]{Chen:2016:XST:2939672.2939785}
Chen, T., \& Guestrin, C. 2016, in Proceedings of the 22Nd ACM SIGKDD
  International Conference on Knowledge Discovery and Data Mining, KDD '16 (New
  York, NY, USA: ACM), 785--794.
\newblock \url{http://doi.acm.org/10.1145/2939672.2939785}

\bibitem[{{Claesen} \& {De Moor}(2015)}]{2015arXiv150202127C}
{Claesen}, M., \& {De Moor}, B. 2015, arXiv e-prints, arXiv:1502.02127

\bibitem[{{Cushing} {et~al.}(2018){Cushing}, {Moskovitz}, \&
  {Gustafsson}}]{2018RNAAS...2b..50C}
{Cushing}, M.~C., {Moskovitz}, N., \& {Gustafsson}, A. 2018, Research Notes of
  the American Astronomical Society, 2, 50

\bibitem[{{Cutri} {et~al.}(2012){Cutri}, {Wright}, {Conrow}, {Bauer},
  {Benford}, {Brandenburg}, {Dailey}, {Eisenhardt}, {Evans}, {Fajardo-Acosta},
  {Fowler}, {Gelino}, {Grillmair}, {Harbut}, {Hoffman}, {Jarrett},
  {Kirkpatrick}, {Leisawitz}, {Liu}, {Mainzer}, {Marsh}, {Masci}, {McCallon},
  {Padgett}, {Ressler}, {Royer}, {Skrutskie}, {Stanford}, {Wyatt}, {Tholen},
  {Tsai}, {Wachter}, {Wheelock}, {Yan}, {Alles}, {Beck}, {Grav}, {Masiero},
  {McCollum}, {McGehee}, {Papin}, \& {Wittman}}]{2012wise.rept....1C}
{Cutri}, R.~M., {Wright}, E.~L., {Conrow}, T., {et~al.} 2012, {Explanatory
  Supplement to the WISE All-Sky Data Release Products}, Tech. rep.

\bibitem[{{Cutri} {et~al.}(2013){Cutri}, {Wright}, {Conrow}, {Fowler},
  {Eisenhardt}, {Grillmair}, {Kirkpatrick}, {Masci}, {McCallon}, {Wheelock},
  {Fajardo-Acosta}, {Yan}, {Benford}, {Harbut}, {Jarrett}, {Lake}, {Leisawitz},
  {Ressler}, {Stanford}, {Tsai}, {Liu}, {Helou}, {Mainzer}, {Gettings},
  {Gonzalez}, {Hoffman}, {Marsh}, {Padgett}, {Skrutskie}, {Beck}, {Papin}, \&
  {Wittman}}]{2013wise.rept....1C}
---. 2013, {Explanatory Supplement to the AllWISE Data Release Products}, Tech.
  rep.

\bibitem[{{Faherty} {et~al.}(2018){Faherty}, {Gagn{\'e}}, {Burgasser},
  {Mamajek}, {Gonzales}, {Bardalez Gagliuffi}, \&
  {Marocco}}]{2018ApJ...868...44F}
{Faherty}, J.~K., {Gagn{\'e}}, J., {Burgasser}, A.~J., {et~al.} 2018, \apj,
  868, 44

\bibitem[{{Faherty} {et~al.}(2016){Faherty}, {Riedel}, {Cruz}, {Gagne},
  {Filippazzo}, {Lambrides}, {Fica}, {Weinberger}, {Thorstensen}, {Tinney},
  {Baldassare}, {Lemonier}, \& {Rice}}]{2016ApJS..225...10F}
{Faherty}, J.~K., {Riedel}, A.~R., {Cruz}, K.~L., {et~al.} 2016, \apjs, 225, 10

\bibitem[{{Gaia Collaboration} {et~al.}(2018){Gaia Collaboration}, {Brown},
  {Vallenari}, {Prusti}, {de Bruijne}, {Babusiaux}, {Bailer-Jones}, {Biermann},
  {Evans}, {Eyer}, \& et~al.}]{2018A&A...616A...1G}
{Gaia Collaboration}, {Brown}, A.~G.~A., {Vallenari}, A., {et~al.} 2018, \aap,
  616, A1

\bibitem[{{Henry} {et~al.}(2018){Henry}, {Jao}, {Winters}, {Dieterich},
  {Finch}, {Ianna}, {Riedel}, {Silverstein}, {Subasavage}, \&
  {Vrijmoet}}]{2018AJ....155..265H}
{Henry}, T.~J., {Jao}, W.-C., {Winters}, J.~G., {et~al.} 2018, \aj, 155, 265

\bibitem[{{Hora} {et~al.}(2012){Hora}, {Marengo}, {Park}, {Wood}, {Hoffmann},
  {Lowrance}, {Carey}, {Surace}, {Krick}, {Glaccum}, {Ingalls}, {Laine},
  {Fazio}, {Ashby}, \& {Wang}}]{2012SPIE.8442E..39H}
{Hora}, J.~L., {Marengo}, M., {Park}, R., {et~al.} 2012, in \procspie, Vol.
  8442, Space Telescopes and Instrumentation 2012: Optical, Infrared, and
  Millimeter Wave, 844239

\bibitem[{{Kirkpatrick} {et~al.}(2011){Kirkpatrick}, {Cushing}, {Gelino},
  {Griffith}, {Skrutskie}, {Marsh}, {Wright}, {Mainzer}, {Eisenhardt},
  {McLean}, {Thompson}, {Bauer}, {Benford}, {Bridge}, {Lake}, {Petty},
  {Stanford}, {Tsai}, {Bailey}, {Beichman}, {Bloom}, {Bochanski}, {Burgasser},
  {Capak}, {Cruz}, {Hinz}, {Kartaltepe}, {Knox}, {Manohar}, {Masters},
  {Morales-Calder{\'o}n}, {Prato}, {Rodigas}, {Salvato}, {Schurr}, {Scoville},
  {Simcoe}, {Stapelfeldt}, {Stern}, {Stock}, \& {Vacca}}]{2011ApJS..197...19K}
{Kirkpatrick}, J.~D., {Cushing}, M.~C., {Gelino}, C.~R., {et~al.} 2011, \apjs,
  197, 19

\bibitem[{{Kirkpatrick} {et~al.}(2012){Kirkpatrick}, {Gelino}, {Cushing},
  {Mace}, {Griffith}, {Skrutskie}, {Marsh}, {Wright}, {Eisenhardt}, {McLean},
  {Mainzer}, {Burgasser}, {Tinney}, {Parker}, \&
  {Salter}}]{2012ApJ...753..156K}
{Kirkpatrick}, J.~D., {Gelino}, C.~R., {Cushing}, M.~C., {et~al.} 2012, \apj,
  753, 156

\bibitem[{{Kirkpatrick} {et~al.}(2014){Kirkpatrick}, {Schneider},
  {Fajardo-Acosta}, {Gelino}, {Mace}, {Wright}, {Logsdon}, {McLean}, {Cushing},
  {Skrutskie}, {Eisenhardt}, {Stern}, {Balokovi{\'c}}, {Burgasser}, {Faherty},
  {Lansbury}, {Rich}, {Skrzypek}, {Fowler}, {Cutri}, {Masci}, {Conrow},
  {Grillmair}, {McCallon}, {Beichman}, \& {Marsh}}]{2014ApJ...783..122K}
{Kirkpatrick}, J.~D., {Schneider}, A., {Fajardo-Acosta}, S., {et~al.} 2014,
  \apj, 783, 122

\bibitem[{{Kirkpatrick} {et~al.}(2019){Kirkpatrick}, {Martin}, {Smart},
  {Cayago}, {Beichman}, {Marocco}, {Gelino}, {Faherty}, {Cushing}, {Schneider},
  {Mace}, {Tinney}, {Wright}, {Lowrance}, {Ingalls}, {Vrba}, {Munn}, {Dahm}, \&
  {McLean}}]{2019ApJS..240...19K}
{Kirkpatrick}, J.~D., {Martin}, E.~C., {Smart}, R.~L., {et~al.} 2019, \apjs,
  240, 19

\bibitem[{{Kuchner} {et~al.}(2017){Kuchner}, {Faherty}, {Schneider}, {Meisner},
  {Filippazzo}, {Gagn{\'e}}, {Trouille}, {Silverberg}, {Castro}, {Fletcher},
  {Mokaev}, \& {Stajic}}]{2017ApJ...841L..19K}
{Kuchner}, M.~J., {Faherty}, J.~K., {Schneider}, A.~C., {et~al.} 2017, \apjl,
  841, L19

\bibitem[{Kuhn \& Johnson(2013)}]{kuhn2013applied}
Kuhn, M., \& Johnson, K. 2013, Applied predictive modeling, Vol.~26 (Springer)

\bibitem[{{Lawrence} {et~al.}(2007){Lawrence}, {Warren}, {Almaini}, {Edge},
  {Hambly}, {Jameson}, {Lucas}, {Casali}, {Adamson}, {Dye}, {Emerson},
  {Foucaud}, {Hewett}, {Hirst}, {Hodgkin}, {Irwin}, {Lodieu}, {McMahon},
  {Simpson}, {Smail}, {Mortlock}, \& {Folger}}]{2007MNRAS.379.1599L}
{Lawrence}, A., {Warren}, S.~J., {Almaini}, O., {et~al.} 2007, \mnras, 379,
  1599

\bibitem[{{Lindegren} {et~al.}(2018){Lindegren}, {Hern{\'a}ndez}, {Bombrun},
  {Klioner}, {Bastian}, {Ramos-Lerate}, {de Torres}, {Steidelm{\"u}ller},
  {Stephenson}, {Hobbs}, {Lammers}, {Biermann}, {Geyer}, {Hilger}, {Michalik},
  {Stampa}, {McMillan}, {Casta{\~n}eda}, {Clotet}, {Comoretto}, {Davidson},
  {Fabricius}, {Gracia}, {Hambly}, {Hutton}, {Mora}, {Portell}, {van Leeuwen},
  {Abbas}, {Abreu}, {Altmann}, {Andrei}, {Anglada}, {Balaguer-N{\'u}{\~n}ez},
  {Barache}, {Becciani}, {Bertone}, {Bianchi}, {Bouquillon}, {Bourda},
  {Br{\"u}semeister}, {Bucciarelli}, {Busonero}, {Buzzi}, {Cancelliere},
  {Carlucci}, {Charlot}, {Cheek}, {Crosta}, {Crowley}, {de Bruijne}, {de
  Felice}, {Drimmel}, {Esquej}, {Fienga}, {Fraile}, {Gai}, {Garralda},
  {Gonz{\'a}lez-Vidal}, {Guerra}, {Hauser}, {Hofmann}, {Holl}, {Jordan},
  {Lattanzi}, {Lenhardt}, {Liao}, {Licata}, {Lister}, {L{\"o}ffler},
  {Marchant}, {Martin-Fleitas}, {Messineo}, {Mignard}, {Morbidelli}, {Poggio},
  {Riva}, {Rowell}, {Salguero}, {Sarasso}, {Sciacca}, {Siddiqui}, {Smart},
  {Spagna}, {Steele}, {Taris}, {Torra}, {van Elteren}, {van Reeven}, \&
  {Vecchiato}}]{2018A&A...616A...2L}
{Lindegren}, L., {Hern{\'a}ndez}, J., {Bombrun}, A., {et~al.} 2018, \aap, 616,
  A2

\bibitem[{{Luhman}(2014)}]{2014ApJ...786L..18L}
{Luhman}, K.~L. 2014, \apjl, 786, L18

\bibitem[{{Mamajek} {et~al.}(2018){Mamajek}, {Marocco}, {Rees}, {Smart},
  {Cooper}, \& {Burgasser}}]{2018RNAAS...2d.205M}
{Mamajek}, E.~E., {Marocco}, F., {Rees}, J.~M., {et~al.} 2018, Research Notes
  of the American Astronomical Society, 2, 205

\bibitem[{{Marsh} \& {Jarrett}(2012)}]{2012PASA...29..269M}
{Marsh}, K.~A., \& {Jarrett}, T.~H. 2012, \pasa, 29, 269

\bibitem[{{Masci}(2013)}]{2013ascl.soft02010M}
{Masci}, F. 2013, {ICORE: Image Co-addition with Optional Resolution
  Enhancement}, Astrophysics Source Code Library, , , ascl:1302.010

\bibitem[{{McMahon} {et~al.}(2013){McMahon}, {Banerji}, {Gonzalez}, {Koposov},
  {Bejar}, {Lodieu}, {Rebolo}, \& {VHS Collaboration}}]{2013Msngr.154...35M}
{McMahon}, R.~G., {Banerji}, M., {Gonzalez}, E., {et~al.} 2013, The Messenger,
  154, 35

\bibitem[{{Meisner} {et~al.}(2018{\natexlab{a}}){Meisner}, {Lang}, \&
  {Schlegel}}]{2018AJ....156...69M}
{Meisner}, A.~M., {Lang}, D., \& {Schlegel}, D.~J. 2018{\natexlab{a}}, \aj,
  156, 69

\bibitem[{{Meisner} {et~al.}(2018{\natexlab{b}}){Meisner}, {Lang}, \&
  {Schlegel}}]{2018RNAAS...2d.202M}
{Meisner}, A.~M., {Lang}, D.~A., \& {Schlegel}, D.~J. 2018{\natexlab{b}},
  Research Notes of the American Astronomical Society, 2, 202

\bibitem[{Pedregosa {et~al.}(2011)Pedregosa, Varoquaux, Gramfort, Michel,
  Thirion, Grisel, Blondel, Prettenhofer, Weiss, Dubourg,
  {et~al.}}]{pedregosa2011scikit}
Pedregosa, F., Varoquaux, G., Gramfort, A., {et~al.} 2011, Journal of machine
  learning research, 12, 2825

\bibitem[{{Sanders} {et~al.}(2007){Sanders}, {Salvato}, {Aussel}, {Ilbert},
  {Scoville}, {Surace}, {Frayer}, {Sheth}, {Helou}, {Brooke}, {Bhattacharya},
  {Yan}, {Kartaltepe}, {Barnes}, {Blain}, {Calzetti}, {Capak}, {Carilli},
  {Carollo}, {Comastri}, {Daddi}, {Ellis}, {Elvis}, {Fall}, {Franceschini},
  {Giavalisco}, {Hasinger}, {Impey}, {Koekemoer}, {Le F{\`e}vre}, {Lilly},
  {Liu}, {McCracken}, {Mobasher}, {Renzini}, {Rich}, {Schinnerer}, {Shopbell},
  {Taniguchi}, {Thompson}, {Urry}, \& {Williams}}]{2007ApJS..172...86S}
{Sanders}, D.~B., {Salvato}, M., {Aussel}, H., {et~al.} 2007, \apjs, 172, 86

\bibitem[{{Scholz} \& {Bell}(2018)}]{2018RNAAS...2a..33S}
{Scholz}, R.-D., \& {Bell}, C.~P.~M. 2018, Research Notes of the American
  Astronomical Society, 2, 33

\bibitem[{{Skrutskie} {et~al.}(2006){Skrutskie}, {Cutri}, {Stiening},
  {Weinberg}, {Schneider}, {Carpenter}, {Beichman}, {Capps}, {Chester},
  {Elias}, {Huchra}, {Liebert}, {Lonsdale}, {Monet}, {Price}, {Seitzer},
  {Jarrett}, {Kirkpatrick}, {Gizis}, {Howard}, {Evans}, {Fowler}, {Fullmer},
  {Hurt}, {Light}, {Kopan}, {Marsh}, {McCallon}, {Tam}, {Van Dyk}, \&
  {Wheelock}}]{2006AJ....131.1163S}
{Skrutskie}, M.~F., {Cutri}, R.~M., {Stiening}, R., {et~al.} 2006, \aj, 131,
  1163

\bibitem[{{Smart} {et~al.}(2019){Smart}, {Marocco}, {Sarro}, {Barrado},
  {Beam{\'{\i}}n}, {Caballero}, \& {Jones}}]{2019MNRAS.tmp..664S}
{Smart}, R.~L., {Marocco}, F., {Sarro}, L.~M., {et~al.} 2019, \mnras,
  arXiv:1902.07571

\bibitem[{Tan(2018)}]{tan2018introduction}
Tan, P.-N. 2018, Introduction to data mining (Pearson Education India)

\bibitem[{{Wang} {et~al.}(2018){Wang}, {Smart}, {Shao}, {Jones}, {Marocco},
  {Luo}, {Burgasser}, {Zhong}, \& {Du}}]{2018PASP..130f4402W}
{Wang}, Y., {Smart}, R.~L., {Shao}, Z., {et~al.} 2018, \pasp, 130, 064402

\bibitem[{{Wright} {et~al.}(2010){Wright}, {Eisenhardt}, {Mainzer}, {Ressler},
  {Cutri}, {Jarrett}, {Kirkpatrick}, {Padgett}, {McMillan}, {Skrutskie},
  {Stanford}, {Cohen}, {Walker}, {Mather}, {Leisawitz}, {Gautier}, {McLean},
  {Benford}, {Lonsdale}, {Blain}, {Mendez}, {Irace}, {Duval}, {Liu}, {Royer},
  {Heinrichsen}, {Howard}, {Shannon}, {Kendall}, {Walsh}, {Larsen}, {Cardon},
  {Schick}, {Schwalm}, {Abid}, {Fabinsky}, {Naes}, \&
  {Tsai}}]{2010AJ....140.1868W}
{Wright}, E.~L., {Eisenhardt}, P.~R.~M., {Mainzer}, A.~K., {et~al.} 2010, \aj,
  140, 1868

\bibitem[{{Wright} {et~al.}(2014){Wright}, {Mainzer}, {Kirkpatrick}, {Masci},
  {Cushing}, {Bauer}, {Fajardo-Acosta}, {Gelino}, {Beichman}, {Skrutskie},
  {Grav}, {Eisenhardt}, \& {Cutri}}]{2014AJ....148...82W}
{Wright}, E.~L., {Mainzer}, A., {Kirkpatrick}, J.~D., {et~al.} 2014, \aj, 148,
  82

\bibitem[{{York} {et~al.}(2000){York}, {Adelman}, {Anderson}, {Anderson},
  {Annis}, {Bahcall}, {Bakken}, {Barkhouser}, {Bastian}, {Berman}, {Boroski},
  {Bracker}, {Briegel}, {Briggs}, {Brinkmann}, {Brunner}, {Burles}, {Carey},
  {Carr}, {Castander}, {Chen}, {Colestock}, {Connolly}, {Crocker}, {Csabai},
  {Czarapata}, {Davis}, {Doi}, {Dombeck}, {Eisenstein}, {Ellman}, {Elms},
  {Evans}, {Fan}, {Federwitz}, {Fiscelli}, {Friedman}, {Frieman}, {Fukugita},
  {Gillespie}, {Gunn}, {Gurbani}, {de Haas}, {Haldeman}, {Harris}, {Hayes},
  {Heckman}, {Hennessy}, {Hindsley}, {Holm}, {Holmgren}, {Huang}, {Hull},
  {Husby}, {Ichikawa}, {Ichikawa}, {Ivezi{\'c}}, {Kent}, {Kim}, {Kinney},
  {Klaene}, {Kleinman}, {Kleinman}, {Knapp}, {Korienek}, {Kron}, {Kunszt},
  {Lamb}, {Lee}, {Leger}, {Limmongkol}, {Lindenmeyer}, {Long}, {Loomis},
  {Loveday}, {Lucinio}, {Lupton}, {MacKinnon}, {Mannery}, {Mantsch}, {Margon},
  {McGehee}, {McKay}, {Meiksin}, {Merelli}, {Monet}, {Munn}, {Narayanan},
  {Nash}, {Neilsen}, {Neswold}, {Newberg}, {Nichol}, {Nicinski}, {Nonino},
  {Okada}, {Okamura}, {Ostriker}, {Owen}, {Pauls}, {Peoples}, {Peterson},
  {Petravick}, {Pier}, {Pope}, {Pordes}, {Prosapio}, {Rechenmacher}, {Quinn},
  {Richards}, {Richmond}, {Rivetta}, {Rockosi}, {Ruthmansdorfer}, {Sandford},
  {Schlegel}, {Schneider}, {Sekiguchi}, {Sergey}, {Shimasaku}, {Siegmund},
  {Smee}, {Smith}, {Snedden}, {Stone}, {Stoughton}, {Strauss}, {Stubbs},
  {SubbaRao}, {Szalay}, {Szapudi}, {Szokoly}, {Thakar}, {Tremonti}, {Tucker},
  {Uomoto}, {Vanden Berk}, {Vogeley}, {Waddell}, {Wang}, {Watanabe},
  {Weinberg}, {Yanny}, \& {Yasuda}}]{2000AJ....120.1579Y}
{York}, D.~G., {Adelman}, J., {Anderson}, Jr., J.~E., {et~al.} 2000, \aj, 120,
  1579

\end{thebibliography}



\end{document}